\documentclass[12pt]{article}
\usepackage{ejorsj-s2}
\usepackage{amsmath,amssymb,amsfonts,amsthm}
\usepackage{graphicx}
\eqswitchon

\usepackage{bm}

\usepackage{color}

\numberwithin{equation}{section}

\newtheorem{proposition}{Proposition}[section]

\numberwithin{remark}{section} \numberwithin{proposition}{section}
\numberwithin{corollary}{section}

\newcommand {\R}{\mathbb{R}}

\newcommand {\p}{\mathbb{P}}

\newcommand {\E}{\mathbb{E}}

\newcommand{\diff}{{\rm d}}

\newcommand{\lev}{L\'{e}vy }

\title{Phase-type Approximation of the Gerber-Shiu Function}
\author{
\begin{tabular}[h]{c}  
Kazutoshi Yamazaki                      \\ 
\textit{Kansai University}\\ 
\end{tabular}
}
\date{(September 09, 2016)}
\begin{document}
\maketitle
\begin{abstract}
The Gerber-Shiu function provides a way of measuring the risk of an
insurance company.  It is given by the expected value of a function
that depends on the ruin time, the deficit at ruin, and the surplus
prior to ruin.  Its computation requires the evaluation of the
overshoot/undershoot distributions of the surplus process at ruin.  In
this paper, we use the recent developments of the fluctuation theory and  approximate it in a closed form by fitting the
underlying process by phase-type \lev processes. A sequence of numerical results are given. 
\end{abstract}

\section{Introduction} \label{section_introduction}

The fundamental objective of the actuarial ruin theory is to measure the vulnerability to insolvency.  Typically, the surplus of an insurance company is modeled by a stochastic process, and ruin occurs at the first time it  goes below a certain threshold.  A most classical and important quantity of interest is the ruin probability, and the \emph{Gerber-Shiu function} is its generalization; it is given as an expected discounted value of the cost function that is dependent on the ruin time,  the deficit at ruin, and the surplus prior to ruin. The evaluation of the Gerber-Shiu function involves that of the overshoot and undershoot distributions at the first down-crossing time that do not admit explicit expressions. Hence its computation is in general a challenging task.

In ruin theory, the surplus process is commonly modeled by a stochastic process with downward jumps. Due to the premiums received from the insured persons, the surplus tends to increase constantly.  On the other hand, it experiences sudden downward jumps due to the insurance payments.  The classical \emph{Cram\'er-Lundberg} model uses a compound Poisson process with downward jumps.  Its generalization called the \emph{Sparre-Andersen} model modifies it by allowing the arrival of the claims to follow a general renewal process. 

In the last decade, significant progress has been made in insurance mathematics and ruin theory, thanks to the development of the theory of \lev processes \cite{Bertoin_1996, Kyprianou_2006}. In particular, many results on the Cram\'er-Lundberg model have been generalized for a general spectrally negative \lev process, or the \lev process with only downward jumps; see, e.g., \cite{Avram_et_al_2007, Bayraktar_2012,Bayraktar_2013, Loeffen_2008}.  This generalization enables one to construct more realistic models; one can, for example, introduce noise by including Brownian motion and/or infinitesimal jumps of infinite activity/variation.   Using the so-called \emph{scale function}, one can express concisely many quantities of interest for a general spectrally negative \lev process.

The objective of this paper is to give an approximation to the Gerber-Shiu function using the theory of scale functions. 
 By the compensation formula of the \lev process, the Gerber-Shiu function admits an expression as a (double) integral with respect to the resolvent measure and the \lev measure.  Because the resolvent can be written using the scale function, at least in theory
 the computation of the Gerber-Shiu function boils down to that of the scale function.

 However, a
major obstacle still remains in putting these in practice because scale functions are in general known
only up to their Laplace transforms, and only a few cases
admit explicit expressions.   The most straightforward approach of computing the scale function is to apply numerical Laplace inversion as in  \cite{Kuznetsov_2011,Surya_2008}.  However, this approach is not suitable for the computation of the Gerber-Shiu function because the numerically approximated scale functions need to be further integrated with respect to the \lev measure.  In particular, the undershoot density that is essential in the computation of the Gerber-Shiu function tends to have a very peculiar form with a possible spike. 
The approximation  hence requires a high precision in computing the scale function and minimal discretization errors in numerical integration.

In this paper, we adopt  a \emph{phase-type fitting} of \cite{Egami_Yamazaki_2010_2} by using the scale function for the class of spectrally negative phase-type \lev processes, or \lev processes with negative phase-type-distributed jumps. Consider a continuous-time Markov chain with some initial distribution and state space consisting of a single absorbing state and a finite number of transient states.  A phase-type distribution is that of the first entry time to the absorbing state.  As has been discussed in \cite{Egami_Yamazaki_2010_2,Kuznetsov_2011}, the scale function of this process becomes the sum of (possibly complex) exponentials; it can be integrated with respect to the \lev measure analytically to obtain an explicit form of the Gerber-Shiu function.  More importantly, the class of phase-type distributions is dense in the class of all positive-valued distributions.
Consequently, the Gerber-Shiu function of any given spectrally negative \lev process can be approximated  \emph{in a closed form} by that of an approximating spectrally negative phase-type \lev process.  Our aim is to evaluate numerically the practicability of this approach.

 In our numerical results, we focus on the case where the \lev measure is finite and has a completely monotone density.  In this case, the jump size distribution can be approximated by a special class of phase-type distributions, called the \emph{hyperexponential}  distributions.  While fitting a phase-type distribution for a general distribution is often difficult, fitting a hyperexponential distribution for the ones with completely monotone densities can be efficiently done, for example by the algorithm by \cite{Feldmann_1998} that is guaranteed to converge to the desired distribution.   The class of \lev processes with completely monotone \lev densities includes, for example, a subset of compound Poisson processes,  variance gamma
\cite{Madan_1998,Madan_1991}, CGMY \cite{CGMY_2002},
generalized hyperbolic \cite{Eberlein_1998} and normal
inverse Gaussian \cite{Barndorff_1998} processes.

 In order to evaluate our approach, we obtain,  for the hyperexponential case, the closed-form expressions of the (discounted) overshoot/undershoot distributions at the first down-crossing time, and use it to approximate those for the processes with Weibull/Pareto-type jumps.  The obtained results are then compared with those obtained by Monte Carlo simulation.
 
 
 To our best knowledge, this is the first paper on the numerical evaluation of the Gerber-Shiu function via phase-type fitting.  As the Gerber-Shiu measure is sensitive to approximation errors, it is important to evaluate its numerical performance.  Recently, the resolvents of related extensions of the \lev process have been developed, and it is reasonable to conjecture that the Gerber-Shiu function of these can be approximated precisely in the same way.

The rest of the paper is organized as follows.  Section \ref{section_model} reviews the spectrally negative \lev process, the Gerber-Shiu function, and the scale function.  Section \ref{scale_phase} gives a summary of \cite{Egami_Yamazaki_2010_2} on the spectrally negative phase-type \lev process and its scale function. Section \ref{section_numerical_results} computes the Gerber-Shiu measure of the hyperexponential \lev process as an approximation for the case the jump size distribution admits a completely monotone density.
We evaluate the performance using numerical results in  Section  \ref{section_numerics}.  Section  \ref{section_concluding_remarks} concludes with remarks on the cases of other variants of spectrally negative \lev processes.

\section{Gerber-Shiu Functions for Spectrally Negative \lev Processes} \label{section_model}
Let $(\Omega,\mathcal{F},\mathbb{P})$ be a probability space hosting a \emph{spectrally negative} \lev process $X = \left\{X_t; t \geq 0 \right\}$ that models the surplus of a company.  Let $\mathbb{P}^x$ be the conditional probability under which $X_0 = x$ (and also $\mathbb{P} \equiv \mathbb{P}^0$), and $\mathbb{F} := \left\{ \mathcal{F}_t: t \geq 0 \right\}$ the filtration generated by $X$.  The process $X$ is uniquely characterized by its \emph{Laplace exponent}
\begin{align}
\psi(s)  := \log \E \left[ e^{s X_1} \right] =  c s +\frac{1}{2}\sigma^2 s^2 + \int_{(-\infty,0)} (e^{s z}-1 - s z 1_{\{z > -1\}}) \Pi (\diff z), \quad s \geq 0, \label{laplace_spectrally_negative}
\end{align}
where $\sigma \geq 0$ is the diffusion (Brownian motion) coefficient and  $\Pi$ is a \lev measure with the support $(-\infty,0)$ that satisfies the integrability condition $\int_{(-\infty,0)} (1 \wedge |z|^2) \Pi(\diff z) < \infty$.  It has paths of bounded variation if and only if
\begin{align*}
\sigma = 0 \quad \textrm{and} \quad \int_{(-\infty,0)} (1 \wedge |z|) \Pi(\diff z) < \infty;
\end{align*}
see, for example, Lemma 2.12 of \cite{Kyprianou_2006}. In this case, we can rewrite the Laplace exponent (\ref{laplace_spectrally_negative}) by
\begin{align*}
\psi(s) = \mu s + \int_{(-\infty,0)} (e^{sz} - 1) \Pi (\diff z), 
\end{align*}
with
\begin{align*}
\mu := c - \int_{(-1,0)} z \Pi (\diff z).
\end{align*}
We disregard the case when $X$ is the negative of a subordinator (or decreasing a.s.).


\subsection{Gerber-Shiu functions} \label{subsection_gerber_shiu}
Define the ruin time as the first time the surplus goes below zero:
\begin{align*}
\tau_0^- := \inf \left\{ t \geq 0: X_t < 0 \right\}.
\end{align*}
Here and throughout the paper, we use the convention that $\inf \emptyset = \infty$.
On the event $\{ \tau_0^-  < \infty \}$, the random variables $X_{\tau_0^-}$ and $X_{\tau_0^- - }$ model, respectively, the deficit at ruin and surplus immediately before ruin. 

Fix $f : (-\infty, 0] \times [0, \infty) \rightarrow [0,\infty)$ bounded and measurable.
We define the \emph{Gerber-Shiu function}
\begin{align*}
GS_{f}(x, q) := \E^x \left[ e^{- q \tau_0^-} f\big( X_{\tau_0^-}, X_{\tau_0^- -} \big); \tau_0^- < \infty \right].
\end{align*}
With the \emph{Gerber-Shiu measure}
\begin{align*}
K^{(q)} (x, \diff y, \diff z) := \E^x \left[e^{- q \tau_0^-}; X_{\tau_0^-} \in \diff y, X_{\tau_0^- -} \in \diff z,\tau_0^- < \infty \right], \quad x, z > 0, \; y < 0,
\end{align*}
we can write 
\begin{align*}
GS_{f}(x, q) = \int_{(0, \infty)} \int_{(-\infty,0)} f(y, z) K^{(q)} (x, \diff y, \diff z);
\end{align*}
see pages 4 and 5 of \cite{kyprianou2013gerber}.

Using the compensation formula (see Theorem 4.4 of \cite{Kyprianou_2006}), this can be written in terms of the  $q$-\emph{resolvent measure} of $X$ killed on exiting $[0,\infty)$:
\begin{align}
R^{(q)} (x, \diff z) := \int_0^\infty e^{-q t} \p^x \{ X_t \in \diff z, \tau_0^- > t \} \diff t,  \quad x, z > 0, \label{resolvent_measure}
\end{align}
which is known to admit a density $ r^{(q)} $ for the case of spectrally negative \lev processes such that
\begin{align}
R^{(q)} (x, \diff z) = r^{(q)} (x, z) \diff z, \quad x, z > 0; \label{resolvent_density}
\end{align}
see \eqref{resolvent} below for the form of $r^{(q)}$.  As in Section 1.3 of \cite{kyprianou2013gerber}, we can write
\begin{align}
K^{(q)} (x, \diff y, \diff z) = \Pi (\diff y - z) r^{(q)} (x, z) \diff z. \label{gerber-shiu_in_terms_of_resolvent}
\end{align}
Hence the computation of the Gerber-Shiu measure boils down to that of the resolvent measure.

\subsection{Scale functions} 
In order to compute the resolvent measure \eqref{resolvent_measure}, we shall introduce the scale function.

Fix $q \geq 0$. The scale function $W^{(q)}: \R \rightarrow [0,\infty)$ of $X$ is a function whose Laplace transform is given by
\begin{align}\label{eq:scale}
\int_0^\infty e^{-s x} W^{(q)}(x) \diff x = \frac 1
{\psi(s)-q}, \qquad s > \Phi(q)
\end{align}
where
\begin{align}
\Phi(q) :=\sup\{s  \geq 0: \psi(s)=q\}, \quad
q\ge 0. \label{zeta}
\end{align}
On the negative half line, it is assumed that $W^{(q)}(x)=0$.


Regarding the smoothness of the scale function, if the \lev measure has no atoms or $X$ is of unbounded variation, then $W^{(q)} \in C^1(0,\infty)$; if it has a Gaussian component ($\sigma > 0$), then $W^{(q)} \in C^2(0,\infty)$.  See \cite{Chan_2009} for other known results on the smoothness.  

The behavior in the neighborhood of zero is given as follows.  As in Lemmas 4.3 and 4.4 of \cite{Kyprianou_Surya_2007},
for every $q \geq 0$, 
\begin{align} \label{asymptotics_zero}
\begin{split}
W^{(q)} (0) &= \left\{ \begin{array}{ll} 0, & \textrm{if $X$ is of unbounded variation} \\ \frac 1 {\mu}, & \textrm{if $X$ is of bounded variation} \end{array} \right\}, \\
W^{(q)'} (0+) &= \left\{ \begin{array}{ll}  \frac 2 {\sigma^2}, & \textrm{if } \sigma > 0 \\   \infty, & \textrm{if } \sigma = 0 \; \textrm{and} \; \Pi(-\infty,0) = \infty \\ \frac {q + \Pi(-\infty,0)} {\mu^2}, & \textrm{if $X$ is compound Poisson} \end{array} \right\}.
\end{split}
\end{align}

\begin{figure}[htbp]
\begin{center}
\begin{minipage}{1.0\textwidth}
\centering
\begin{tabular}{cc}
\includegraphics[scale=0.58]{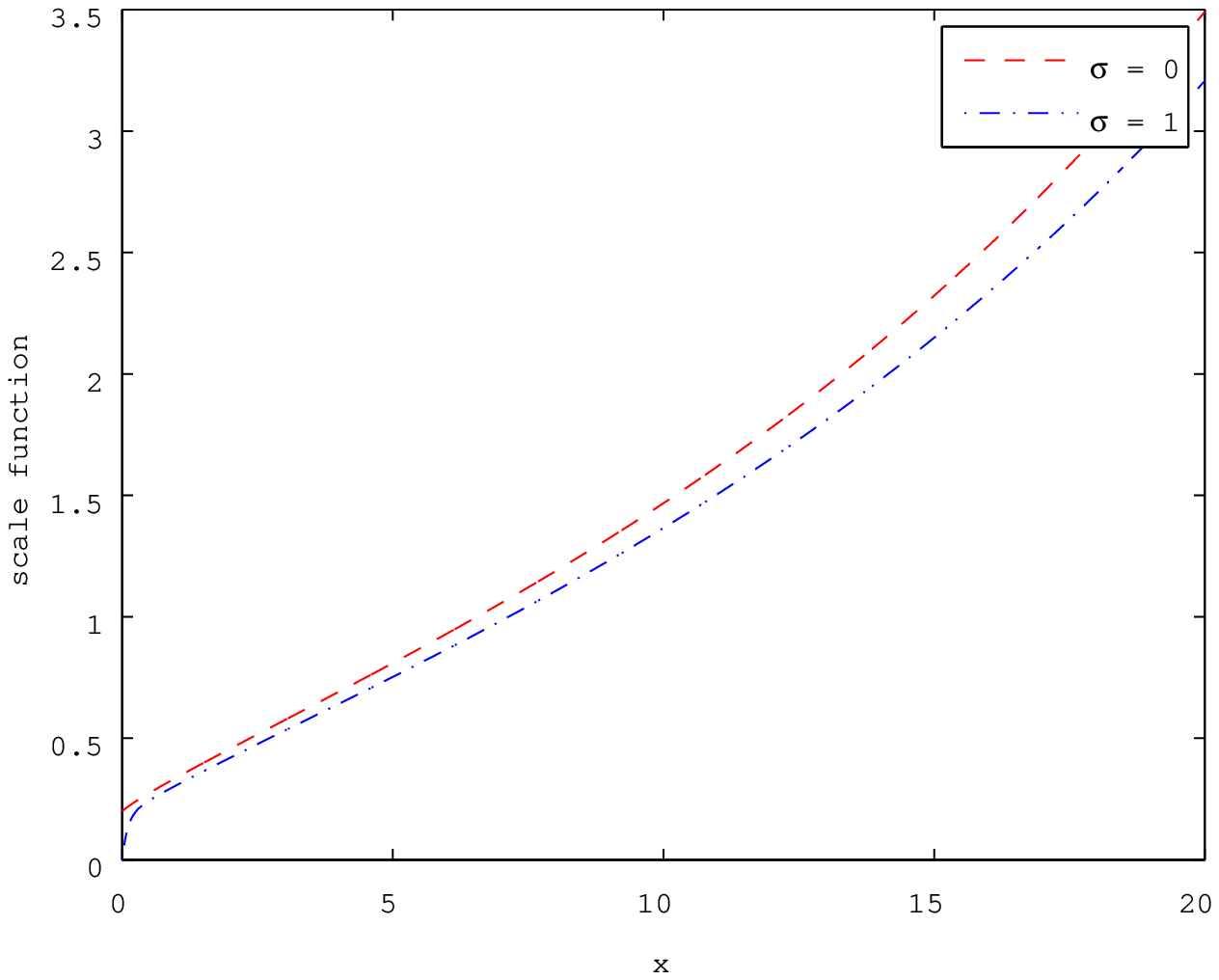}  & \includegraphics[scale=0.58]{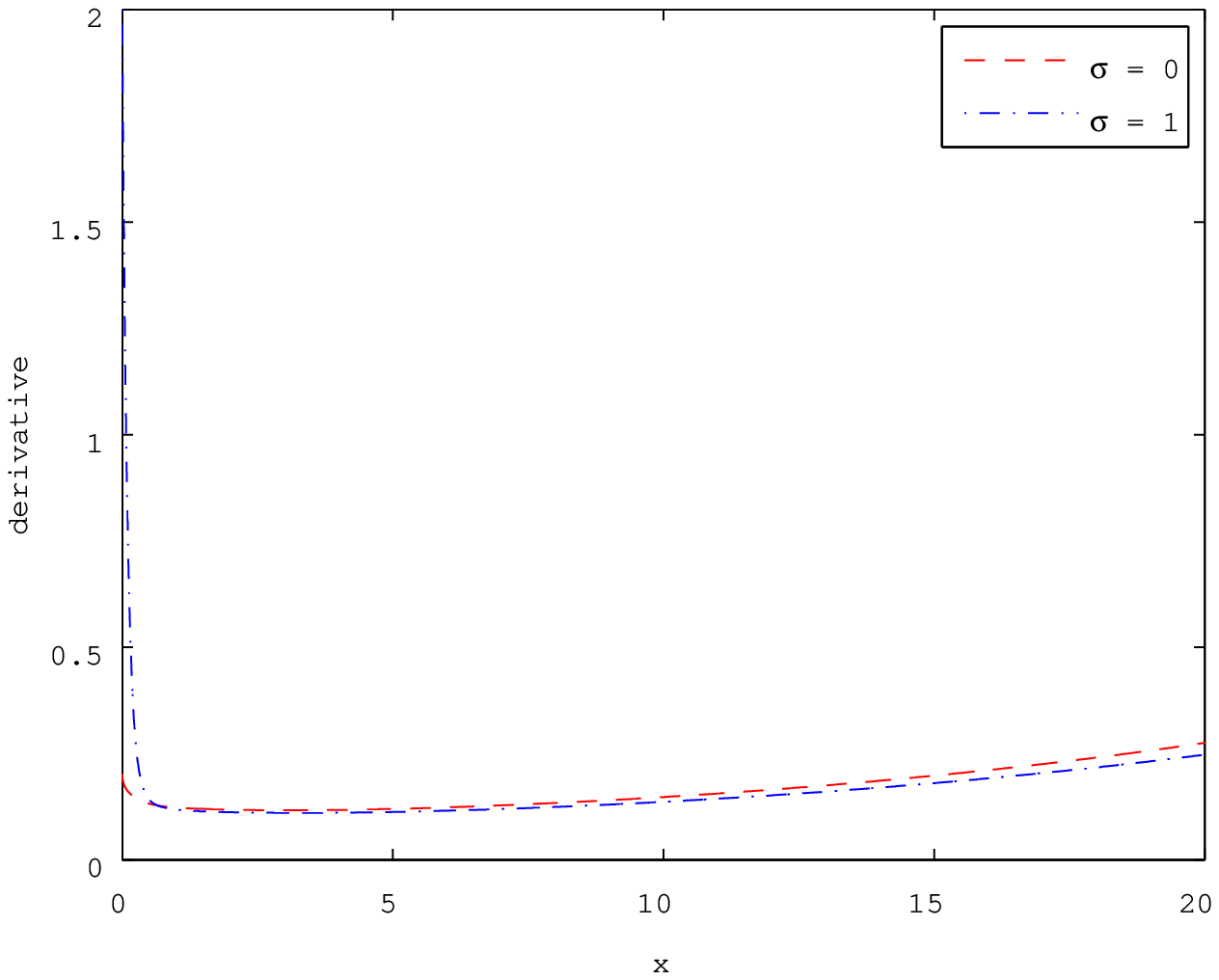} 
\end{tabular}
\end{minipage}
\caption{Sample plots of the scale function (left) and its derivative (right).  The ones indicated in red (resp.\ blue) are those for the case of bounded (resp.\ unbounded) variation.  As in \eqref{asymptotics_zero}, it is confirmed that it vanishes at zero if and only if it is of unbounded variation.}
\label{fig:weibull_scale}
\end{center}
\end{figure}

We plot in Figure \ref{fig:weibull_scale} sample plots of the scale function and its derivative for the bounded variation case with $\sigma = 0$ and the unbounded variation case with $\sigma > 0$.

The most well-known application of the scale function can be found in the two-sided exit identities. Let us define the first down- and up-crossing times, respectively, of $X$ by
\begin{align}
\label{first_passage_time}
\tau_b^- := \inf \left\{ t > 0: X_t < b \right\} \quad \textrm{and} \quad \tau_b^+ := \inf \left\{ t > 0: X_t >  b \right\}, \quad b \in \R.
\end{align}
Then, for any $b > 0$ and $x \leq b$,
\begin{align}
\begin{split}
\E^x \left[ e^{-q \tau_b^+} 1_{\left\{ \tau_b^+ < \tau_0^- \right\}}\right] &= \frac {W^{(q)}(x)}  {W^{(q)}(b)}, \\
 \E^x \left[ e^{-q \tau_0^-} 1_{\left\{ \tau_b^+ > \tau_0^- \right\}}\right] &= Z^{(q)}(x) -  Z^{(q)}(b) \frac {W^{(q)}(x)}  {W^{(q)}(b)}, \\
 \E^x \left[ e^{-q \tau_0^-} \right] &= Z^{(q)}(x) -  \frac q {\Phi(q)} W^{(q)}(x),
\end{split}
 \label{laplace_in_terms_of_z}
\end{align}
where
\begin{align*}
Z^{(q)}(x) &:= 1 + q \int_0^x W^{(q)}(y) \diff y, \quad x \in \R.
\end{align*}
For a comprehensive account on the scale function, see among others \cite{Kuznetsov_2011, Kyprianou_2006}.

\subsection{Resolvent via the scale function} Recall, as in our discussion in Section \ref{subsection_gerber_shiu}, that the Gerber-Shiu function can be written in terms of the resolvent measure. This can be concisely written in terms of the scale function as follows: by Corollary 8.8 of \cite{Kyprianou_2006}, the resolvent density \eqref{resolvent_density} can be written
\begin{align}
r^{(q)} (x, z) = e^{- \Phi(q) z} W^{(q)} (z) - W^{(q)} (x-z), \quad x, z > 0. \label{resolvent}
\end{align}
Now in view of the identity \eqref{gerber-shiu_in_terms_of_resolvent}, the computation of the Gerber-Shiu measure boils down to that of the scale function.  

\section{Scale Functions for Spectrally Negative Phase-type \lev Processes}
\label{scale_phase}

As discussed in the previous section, the scale function is defined by its Laplace transform and for its computation the Laplace transform \eqref{eq:scale} must be inverted.  Here, we review the results on \cite{Egami_Yamazaki_2010_2} for a special class of \lev processes where it can be inverted analytically.

\subsection{Phase-type distribution}
Consider a continuous-time Markov chain 
\begin{align*}
Y = \{ Y_t; t \geq 0 \}
\end{align*}
 with a finite
state space $\{1,\ldots,m \} \cup \{ \Delta \}$ where $1,\ldots,m$
are transient and $\Delta$ is absorbing. Its initial distribution is given by
a simplex ${\bm \alpha}=[\alpha_1, \ldots, \alpha_m]$
such that $\alpha_i=\p \left\{ Y_0=i \right\}$ for every $i = 1,\ldots,m$.  The intensity matrix ${\bm Q}$ is partitioned into the $m$ transient
states and the absorbing state $\Delta$, and is given by
\begin{align*}
{\bm Q}  := \begin{bmatrix} {\bm T} & {\bm t} \\ {\bm 0} & 0 \end{bmatrix}.
\end{align*}
Here  ${\bm T}$  is an  $m \times m$-matrix called the phase-type-generator, and ${\bm t} = - {\bm T} {\bm 1}$ where  ${\bm 1} =
[1,\ldots,1]'$. A distribution is called \emph{phase-type} with
representation $(m, {\bm \alpha}, {\bm T})$ if it is the
distribution of the absorption time to $\Delta$ in the Markov chain
described above. It is known that ${\bm T}$ is non-singular and thus
invertible; see \cite{Asmussen_1996}.  Its distribution and density functions
are given, respectively, by 
\begin{align*}
F(z; \bm \alpha, \bm T) =  1-{\bm \alpha} e^{{\bm T} z} {\bm 1} \quad \textrm{and} \quad f(z; \bm \alpha, \bm T) = {\bm \alpha} e^{{\bm T} z} {\bm t}, \quad z > 0.
\end{align*}

\subsection{Phase-type \lev processes}
Let $X = \left\{X_t; t \geq 0 \right\}$ be a spectrally negative \lev process of the form
\begin{equation}
  X_t  - X_0=\mu t+\sigma B_t - \sum_{n=1}^{N_t} Z_n, \quad 0\le t <\infty, \label{levy_canonical}
\end{equation}
for some $\mu \in \R$ and $\sigma \geq 0$ (with $\mu > 0$ when $\sigma = 0$ so that it is not the negative of a subordinator).  Here $B=\{B_t; t\ge 0\}$ is a standard Brownian motion, $N=\{N_t; t\ge 0\}$ is a Poisson process with arrival rate $\lambda$, and  $Z = \left\{ Z_n; n = 1,2,\ldots \right\}$ is an i.i.d.\ sequence of phase-type-distributed random variables with representation $(m,{\bm \alpha},{\bm T})$. These processes are assumed mutually independent. Its Laplace exponent is then
\begin{align}
 \psi(s)   = \mu s + \frac 1 2 \sigma^2 s^2 + \lambda \left( {\bm \alpha} (s {\bm I} - {\bm{T}})^{-1} {\bm t} -1 \right), \label{laplace_exponent_ph}
 \end{align}
which is analytic for every $s \in \mathbb{C}$ except at the eigenvalues of ${\bm T}$.  

Let $\mathcal{I}_q$ be the set of (the sign-changed) \emph{negative roots}:
\begin{align}
\mathcal{I}_q &:= \left\{ i: \psi (-\xi_{i,q}) = q \; \textrm{and} \; \mathcal{R} (\xi_{i,q}) > 0\right\}, \label{def_I_q}
\end{align}
where $\mathcal{R}(z)$ is the real part of $z \in \mathbb{C}$.
Let
$n$ denote the number of different roots in $\mathcal{I}_q$ and
$m_i$ the multiplicity of a root $\xi_{i,q}$ for $i = 1,\ldots,n$. 

Because the Laplace exponent \eqref{laplace_exponent_ph} has a rational form, it can be inverted analytically by partial fraction decomposition. Hence, in view of \eqref{eq:scale}, the scale function can be obtained.
\begin{proposition}[Section 5.4 of \cite{Kuznetsov_2011} and Proposition 2.1 of \cite{Egami_Yamazaki_2010_2}] \label{proposition_main}
Suppose $q \geq 0$ and $\psi'(0+) < 0$ if $q=0$. Then the scale function is written
\begin{align}
W^{(q)}(x) =  \frac {e^{\Phi(q) x}} {\psi'(\Phi(q))}  - \sum_{i = 1}^n \sum_{k=1}^{m_i} C_{i,q}^{(k)} \frac {x^{k-1}} {(k-1)!} e^{-\xi_{i,q} x}, \quad x \geq 0, \label{scale_function}
\end{align}
where
\begin{align*}
C_{i,q}^{(k)} &:= \left. \frac 1 {(m_i-k)!} \frac {\partial^{m_i-k}} {\partial s^{m_i-k}} \frac { (s+\xi_{i,q})^{m_i}} {q-\psi(s)} \right|_{s = -\xi_{i,q}}, \quad 1 \leq k \leq m_i \; \textrm{and} \; 1 \leq i \leq n.
\end{align*}
In particular, if all the roots in $\mathcal{I}_q$ are distinct, then
\begin{align}
W^{(q)}(x) =  \frac {e^{\Phi(q) x}} {\psi'(\Phi(q))}  - \sum_{i = 1}^n  C_{i,q} e^{-\xi_{i,q} x}, \quad x \geq 0, \label{scale_function_distinct}
\end{align}
where
\begin{align*}
C_{i,q} &:= \left. \frac { s+\xi_{i,q}} {q-\psi(s)} \right|_{s = -\xi_{i,q}} = - \frac 1 {\psi'(-\xi_{i,q})}.
\end{align*}
\end{proposition}

\subsection{Approximation results}
It is known that the class of phase-type distributions is dense in the class of all positive-valued distributions.  Using this, 
Proposition 1 of \cite{Asmussen_2004} shows that, there exists,
for any spectrally negative \lev process $X$, a sequence of
spectrally negative phase-type \lev processes  $X^{(n)}$
converging to $X$ in $D[0,\infty)$.  In other words, $X_1^{(n)} \rightarrow X_1$ in distribution by Corollary VII 3.6 of \cite{Jacod_Shirayev_2003}; see also \cite{Pistorius_2006}.  

Using these results, Egami and Yamazaki \cite{Egami_Yamazaki_2010_2} study the convergence of the corresponding scale function, and show that the approximation is in most cases very accurate.  On the other hand, as they also point out, no existing algorithm is guaranteed to construct a converging sequences and fitting phase-type distributions can get difficult; see, e.g., the case of fitting for the uniform distributed jump size in  \cite{Egami_Yamazaki_2010_2}.  

On the other hand, as we shall discuss next, it is guaranteed to work for the case the jump size admits a completely monotone density.

\subsection{Hyperexponential case}\label{example_hyperexponential} 
As an important example where all the roots in $\mathcal{I}_q$ are distinct and real, we consider the case where $Z$ has a hyperexponential distribution with a density function
\begin{align*}
f (z)  = \sum_{j=1}^m \alpha_j \eta_j e^{- \eta_j z}, \quad z > 0,
\end{align*}
for some $0 < \eta_1 < \cdots < \eta_m < \infty$ and $\alpha_j > 0$ for $1 \leq j \leq m$ such that $\alpha_1 + \cdots + \alpha_m = 1$; this is the phase-type distribution with its Markov chain such that the $m$ transient states are connected only to $\Delta$.  Its Laplace exponent (\ref{laplace_spectrally_negative})
is then
\begin{align*}
\psi(s) = \mu s + \frac 1 2 \sigma^2 s^2  - \lambda
\sum_{j=1}^m \alpha_j \frac s {\eta_j + s}.
\end{align*}
Notice in this case that $-\eta_1$, \ldots, $-\eta_m$ are the poles of the Laplace exponent.  Furthermore, all the roots in $\mathcal{I}_q$ are distinct and real and satisfy the following interlacing condition for every $q > 0$:
\begin{enumerate}
\item
for $\sigma > 0$, there are $m+1$ roots $-\xi_{1,q}, \ldots, -\xi_{m+1,q}$ such that
\begin{align}
0 < \xi_{1,q} < \eta_1 < \xi_{2,q} < \cdots < \eta_m < \xi_{m+1,q} < \infty; \label{interlacing1}
\end{align}
\item for $\sigma =0$, there are $m$ roots $-\xi_{1,q}, \ldots, -\xi_{m,q}$  such that
\begin{align}
0 < \xi_{1,q} < \eta_1 < \xi_{2,q} < \cdots <  \xi_{m,q}  < \eta_m < \infty. \label{interlacing2}
\end{align}
\end{enumerate}
Because all roots are real and distinct, the scale function can be written as \eqref{scale_function_distinct}.

Recall that a density function $f$ of a positive valued random variable is called completely monotone if all the derivatives exist and, for every $n \geq 1$,
\begin{align*}
(-1)^n f^{(n)} (x) \geq 0, \quad x \geq 0,
\end{align*}
where $f^{(n)}$ denotes the $n^{th}$ derivative of $f$. 

Feldmann and Whitt \cite{Feldmann_1998} showed that if a density function is \emph{completely monotone},
then it can be approximated by those  of hyperexponential distributions. 
As shown by \cite{Bernstein_1929}, every completely monotone density function is a mixture of exponential density functions, and this implies that, for any distribution with a completely monotone density, there exists a sequence of hyperexponential distributions converging to it. The class of distributions with completely monotone densities contains a number of distributions such as the (subset of) Pareto distribution, the Weibull distribution, and the gamma distribution.  
Feldmann and Whitt \cite{Feldmann_1998} took advantage of this fact and  proposed a recursive algorithm for fitting hyperexponential distributions to these distributions. We refer the reader to \cite{Albrecher_2010, Kammler_1976} for other approximation methods.

\section{Computation of the Gerber-Shiu Function} \label{section_numerical_results}

We shall now consider the approximation of the Gerber-Shiu function using the fitted scale functions.  Here, we consider the hyperexponential case as discussed in Section \ref{example_hyperexponential}; in this case, because the scale function and \lev measure are both written as mixtures of exponential functions,  we attain closed-form expressions.

\subsection{Overshoot and undershoot distributions} 
Recall from \eqref{gerber-shiu_in_terms_of_resolvent} regarding the equivalence of the Gerber-Shiu measure and the product of the \lev measure and the resolvent measure.  If we define for all $A \in \mathcal{B}(-\infty,0)$ and $B \in \mathcal{B}(0,\infty)$,
\begin{align} 
h_q(x; A, B) := \E^x \left[ e^{-q\tau_0^-};   \, X_{\tau_0^-} \in A, \, X_{\tau_0^--} \in B, \, \tau_0^- < \infty \right] = \int_{A \times B}K^{(q)} (x, \diff y, \diff z). \label{h_q}
\end{align}
Combining \eqref{gerber-shiu_in_terms_of_resolvent} and \eqref{resolvent}, we can write
\begin{align*}
h_q(x; A, B)
= \int_0^\infty   \left\{ W^{(q)}(x)  \int_{B \cap (A+u)} e^{-\Phi(q) y} \diff y  -  \int_{B \cap (A+u)}  W^{(q)}(x-y) \diff y \right\} \overline{\Pi} (\diff u),
\end{align*}
where $\overline{\Pi}$ is the \lev measure of the dual process $-X$.

When $X$ is a phase-type \lev process, as we have studied in the previous section $W^{(q)}(x)$ can be written as a sum of (possibly complex) exponentials.  In particular, if it is hyperexpontial, we can write $\overline{\Pi} (\diff u) = \lambda \sum_{j=1}^m \alpha_j \eta_j e^{-\eta_j u}\diff u$ for all $u \in (0,\infty)$, and hence $h_q$ can be obtained analytically.

Here we assume that $X$ is a hyperexponential \lev process and $q > 0$.  The following results are immediate by straightforward integration.
\begin{proposition} \label{proposition_overshoot_undershoot_hyperexponential}
\begin{enumerate}
\item Suppose $B = (\underline{b}, \overline{b})$ and $A = (-\overline{a}, -\underline{a})$ for some $0 \leq \underline{a} \leq \overline{a}$ and $0 \leq \underline{b} \leq \overline{b}$. Then
\begin{align*}
h_q(x; A, B)
&= \lambda \sum_{j=1}^m \alpha_j  (e^{-\eta_j \underline{a}} - e^{-\eta_j \overline{a}})  \kappa_{j,q} (x; B)
\end{align*}
where, for each $1 \leq j \leq m$,
\begin{multline*}
\kappa_{j,q} (x;B) := \frac {e^{\Phi(q) x}} {\psi'(\Phi(q))(\eta_j+\Phi(q))} \left(  e^{-(\eta_j+\Phi(q)) (\underline{b} \vee x)} - e^{-(\eta_j+\Phi(q)) (\overline{b}\vee x) } \right) \\ + \sum_{i \in \mathcal{I}_q} C_{i,q}e^{-\xi_{i,q} x}  \left[ \frac 1 {\eta_j-\xi_{i,q}} \left(  e^{-(\eta_j-\xi_{i,q}) (\underline{b} \wedge x)} - e^{-(\eta_j-\xi_{i,q}) (\overline{b} \wedge x)} \right) \right. \\ \left.  - \frac 1 {\eta_j+\Phi(q)} \left(  e^{-(\eta_j+\Phi(q)) \underline{b}} - e^{-(\eta_j+\Phi(q)) \overline{b}} \right) \right].
\end{multline*}
\item We have
\begin{multline*}
\begin{aligned}
\E^x \left[ e^{-q\tau_0^-}; -X_{\tau_0^-} \in \diff a, \, X_{\tau_0^--} \in B, \, \tau_0^- < \infty \right] 
&= \lambda \sum_{j=1}^m \alpha_j \eta_j  e^{-\eta_j a}   \kappa_{j,q} (x; B) \\
\E^x \left[ e^{-q\tau_0^-};  X_{\tau_0^-} \in A, \, X_{\tau_0^--} \in \diff b,  \, \tau_0^- < \infty \right] 
&= \lambda \sum_{j=1}^m \alpha_j  (e^{-\eta_j \underline{a}} - e^{-\eta_j \overline{a}})  
\end{aligned}
\\ \times \left\{ \begin{array}{ll}  \sum_{i \in \mathcal{I}_q} C_{i,q}  e^{-\xi_{i,q} x}   \left(   e^{-(\eta_j-\xi_{i,q}) b}  - e^{-(\eta_j+\Phi(q)) b} \right), & b < x\\ \frac 1 {\psi'(\Phi(q))}  e^{\Phi(q) x-(\eta_j+\Phi(q)) b}- \sum_{i \in \mathcal{I}_q} C_{i,q}     e^{-(\xi_{i,q} x+(\eta_j+\Phi(q)) b)},  & b \geq x \end{array} \right\}.
\end{multline*}
In particular, by setting $B = (0,\infty)$ ($A = (-\infty,0)$), 
\begin{multline}
\begin{aligned}
\E^x \left[ e^{-q\tau_0^-}; -X_{\tau_0^-} \in \diff a, \, \tau_0^- < \infty \right] 
&= \lambda \sum_{j=1}^m \alpha_j \eta_j  e^{-\eta_j a}   \kappa_{j,q} (x; (0,\infty)), \\
\E^x \left[ e^{-q\tau_0^-}; X_{\tau_0^--} \in \diff b, \, \tau_0^- < \infty \right]  \end{aligned} \\
=  \left\{ \begin{array}{ll}  \lambda \sum_{j=1}^m \alpha_j  \sum_{i \in \mathcal{I}_q} C_{i,q}  e^{-\xi_{i,q} x}   \left(   e^{-(\eta_j-\xi_{i,q}) b}  - e^{-(\eta_j+\Phi(q)) b} \right), & b < x,\\ \lambda \sum_{j=1}^m \alpha_j  \left[ \frac 1 {\psi'(\Phi(q))}  e^{\Phi(q) x-(\eta_j+\Phi(q)) b}  - \sum_{i \in \mathcal{I}_q} C_{i,q}      e^{-(\xi_{i,q} x+ (\eta_j+\Phi(q)) b)} \right],  & b \geq x, \end{array} \right. 
\label{overshoot_undershoot_density}
\end{multline}
where
\begin{align*}
\kappa_{j,q} (x; (0,\infty)) &=  \frac 1 {\psi'(\Phi(q))(\eta_j+\Phi(q))}  e^{-\eta_j x}  \\ &+ \sum_{i \in \mathcal{I}_q} C_{i,q} \left[ \frac 1 {\eta_j-\xi_{i,q}} \left(  e^{-\xi_{i,q} x} - e^{-\eta_j x} \right) -   e^{-\xi_{i,q} x}  \frac 1 {\eta_j+\Phi(q)} \right].
\end{align*}
\end{enumerate}
\end{proposition}

\section{Numerical Results} \label{section_numerics}
Using the identities obtained in Proposition \ref{proposition_overshoot_undershoot_hyperexponential}, we shall evaluate the efficiency of the phase-type fitting approach of the Gerber-Shiu function.

Here we consider the spectrally negative \lev processes $X^{(\textrm{weibull})}$ and $X^{(\textrm{pareto})}$ in the form (\ref{levy_canonical}) where $Z$ is 
\begin{enumerate}
\item[(i)] Weibull$(0.6,0.665)$, and 
\item[(ii)] Pareto$(1.2,5)$
\end{enumerate}
 respectively. Recall that
the Weibull distribution with parameters $c$ and $a$ (denoted Weibull$(c,a)$) is give by $F(t) = 1-e^{-(t/a)^c}$, $t \geq 0$,
and the Pareto distribution with positive parameters $a$ and $b$ (denoted Pareto$(a,b)$) is given by $F(t) = 1-(1+bt)^{-a}$, $t \geq 0$. See \cite{Johnson_Kotz_1972} for more details about these distributions. With the choice of our parameters, 
the corresponding \lev densities are thus completely monotone.

As has been noted in Section \ref{example_hyperexponential}, any spectrally negative \lev process with a completely monotone \lev density can be approximated arbitrarily closely by fitting  hyperexponential distributions.  Here, we use the fitted data computed by \cite{Feldmann_1998} to approximate the scale function for $X^{(\textrm{weibull})}$ and $X^{(\textrm{pareto})}$ (with or without a Brownian motion component).   Tables 3 and 9, respectively, of \cite{Feldmann_1998} show the parameters of the
hyperexponential distributions obtained by
\cite{Feldmann_1998} fitted to (i) with $m=6$ and to (ii) with
$m=14$.  We use these parameters to construct hyperexponential \lev processes $\widetilde{X}^{(\textrm{weibull})}$ and $\widetilde{X}^{(\textrm{pareto})}$ (see Section \ref{example_hyperexponential}) that will be used to approximate $X^{(\textrm{weibull})}$ and $X^{(\textrm{pareto})}$, respectively.

\begin{table}[ht]
\caption{\small Parameters of the hyperexponential distribution fitted to Weibull($0.6$,$0.665$) and Pareto(1.2,5) (taken from Tables 3 of \cite{Feldmann_1998}).} \label{table:fitted}
\begin{center}
\begin{tabular}{cc}
\centering 
\begin{tabular}{c c c} 
\hline\hline 
$i$ & $\alpha_i$ & $\eta_i$ \\ [0.5ex] 
\hline 
1 & 0.029931 & 676.178 \\ 
2 & 0.093283 & 38.7090 \\
3 & 0.332195 & 4.27400 \\
4 & 0.476233 & 0.76100 \\
5 & 0.068340 & 0.24800 \\
6 & 0.000018 & 0.09700 \\ -- & -- & --
 \\ [1ex]
\hline 
\end{tabular} 
&
\begin{tabular}{c c c | c c c } 
\hline\hline 
$i$ & $\alpha_i$ & $\eta_i$ & $i$ & $\alpha_i$ & $\eta_i$\\ [0.5ex] 
\hline 
1 & 8.37E-11 & 8.3E-09 & 8 & 0.000147 & 0.0020  \\ 
2 & 7.18E-10 & 6.8E-08 & 9 & 0.001122 & 0.0100 \\
3 & 5.56E-09 & 3.9E-07  & 10 & 0.008462 & 0.0570 \\
4 & 4.27E-08 & 2.2E-06& 11 & 0.059768 & 0.3060 \\
5 & 3.27E-07 & 1.2E-05  & 12 & 0.307218 & 1.5460 \\
6 & 2.50E-06 & 6.5E-05  & 13 & 0.533823 & 6.5160 \\
7 & 1.92E-05 & 3.5E-04  & 14 & 0.089437 & 23.304 \\ [1ex] 
\hline 
\end{tabular} \\
(i) Weibull$(0.6,0.665)$  
& \hspace{1cm} (ii) Pareto$(1.2,5)$
\end{tabular}
\end{center}
\end{table}

In order to evaluate the errors associated with the phase-type-fitting of the Gerber-Shiu function, we shall consider approximating the overshoot/undershoot density
\begin{align*}
\E^x \left[ e^{-q \tau_0^-}; -X_{\tau_0^-} \in \diff a, \, \tau_0^- < \infty \right] \quad \textrm{and} \quad \E^x \left[ e^{-q \tau_0^-}; X_{\tau_0^--} \in \diff b, \, \tau_0^- < \infty \right], \end{align*} 
for $X^{(\textrm{weibull})}$  and $X^{(\textrm{pareto})}$. The phase-type-fitting approach approximates them by computing those for the approximating hyperexponential \lev processes $\widetilde{X}^{(\textrm{weibull})}$ and $\widetilde{X}^{(\textrm{pareto})}$.  These can be done analytically by the identity \eqref{overshoot_undershoot_density}.  

We evaluate these results by comparing with the simulated results. 
For $X^{(\textrm{weibull})}$ and $X^{(\textrm{pareto})}$, we simulate
\begin{align*}
&\E^x \left[ e^{-q \tau_0^-}; -X_{\tau_0^-} \in (a-\Delta a/2, a + \Delta a/2), \, \tau_0^- < \infty \right] / \Delta a, \quad \textrm{and} \\ &\E^x \left[ e^{-q \tau_0^-}; X_{\tau_0^--} \in (b-\Delta b/2, b + \Delta b/2), \, \tau_0^- < \infty \right] / \Delta b
\end{align*}
with $\Delta a = \Delta b = 0.1$ by Monte Carlo simulation with $500,000$ samples.  

In order to confirm the accuracy of the simulated results, we also compute the results using these two methods for  $X^{(\textrm{exp})}$ corresponding to the process where the jump size is exponential with parameter $1$; this is a special case of the phase-type (and hyperexponential) \lev process and hence obtained  results are exact.

\begin{figure}[htbp]
\begin{center}
\begin{minipage}{1.0\textwidth}
\centering
\begin{tabular}{cc}
\includegraphics[scale=0.55]{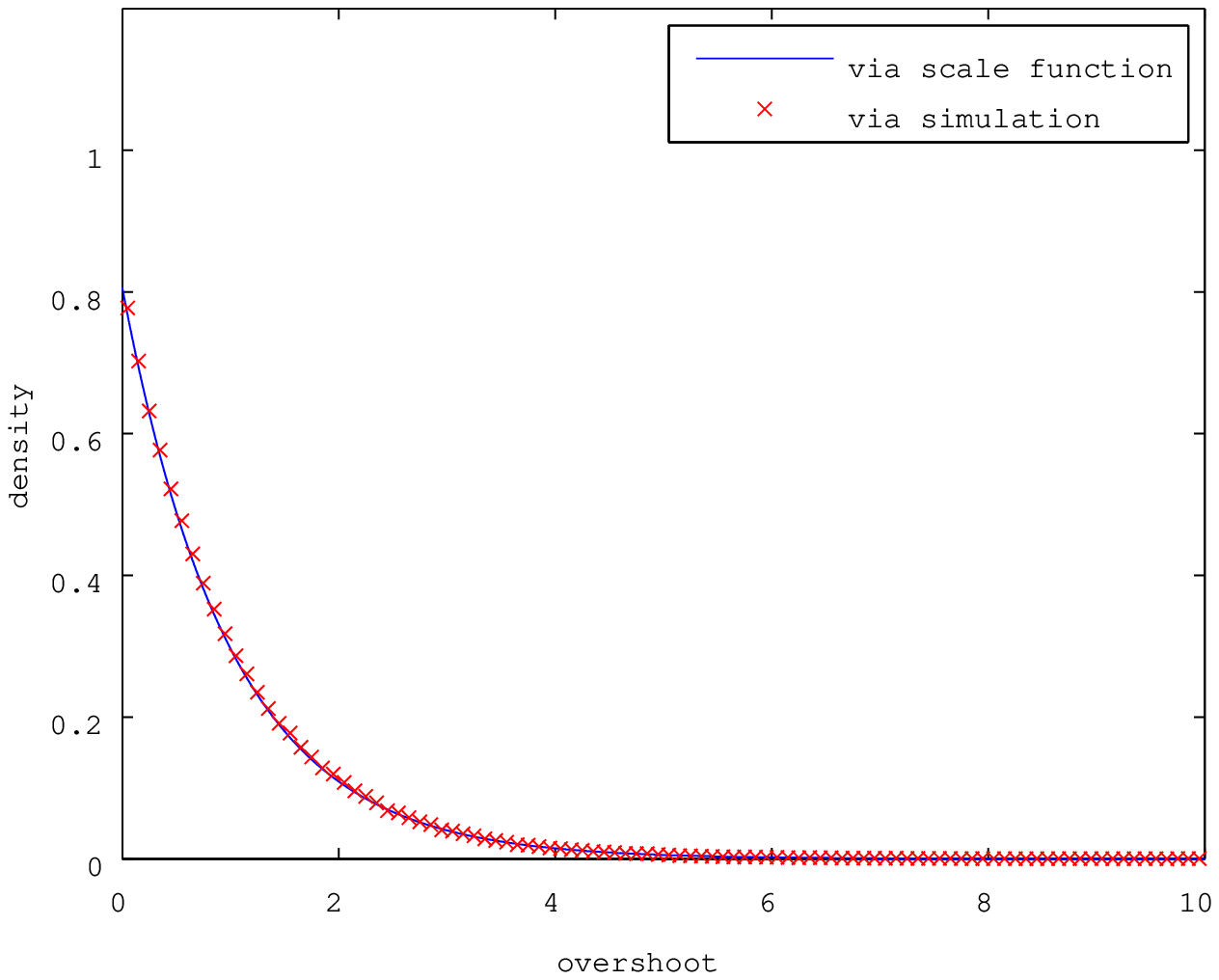}  & \includegraphics[scale=0.55]{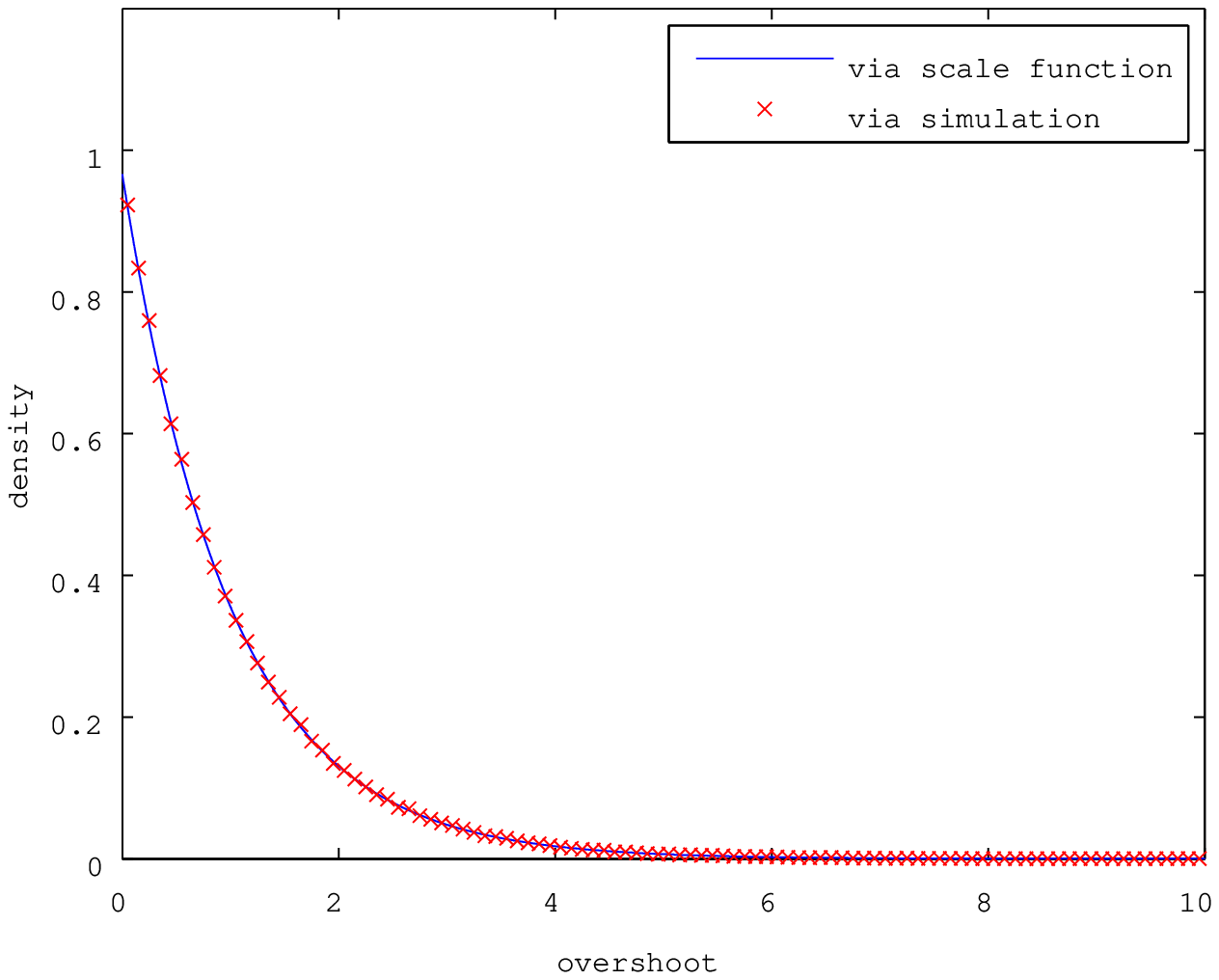} \\
Exp(1)  with $\sigma =1$ & Exp(1)  with $\sigma =0$ \vspace{0.3cm} \\
\includegraphics[scale=0.55]{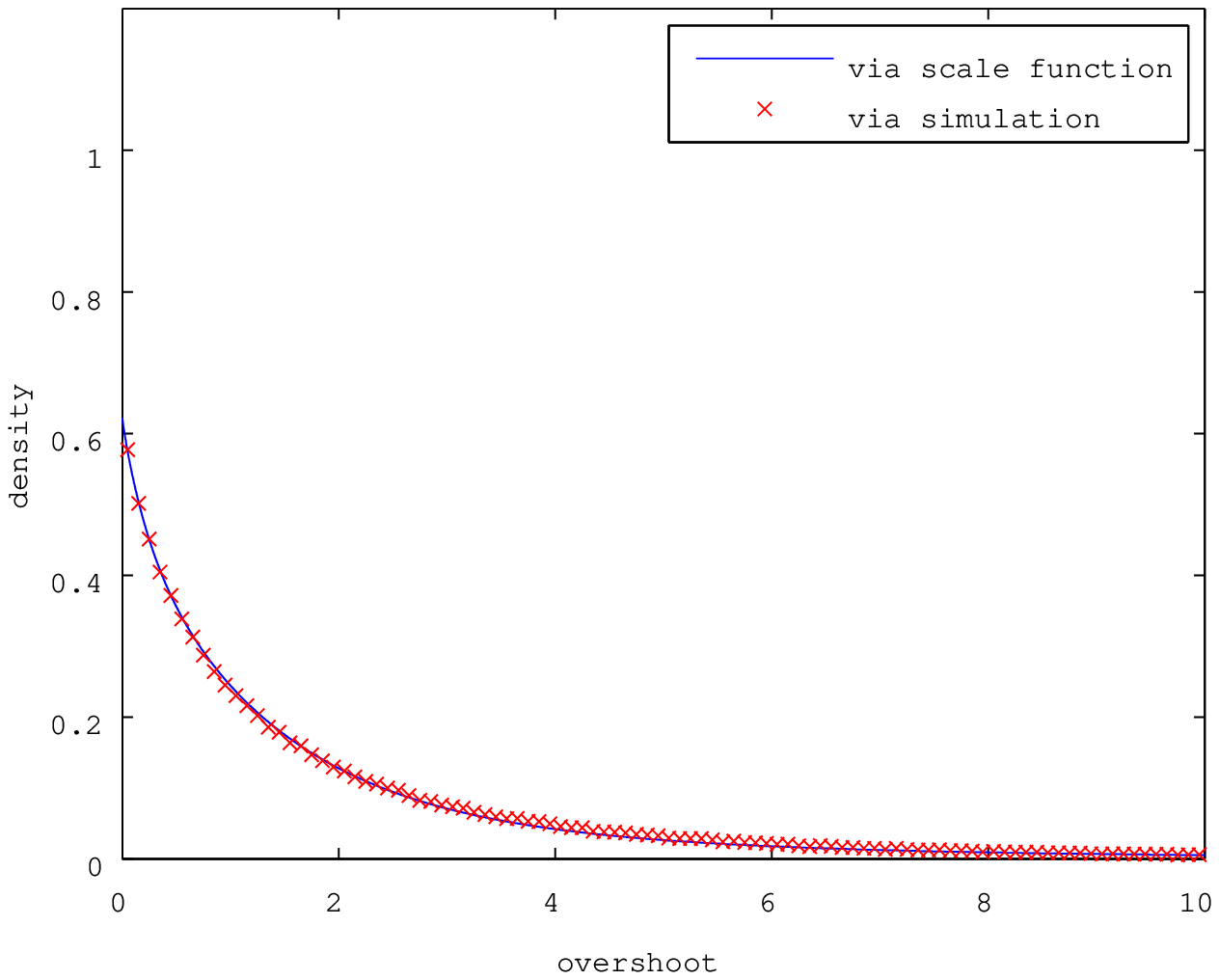}  & \includegraphics[scale=0.55]{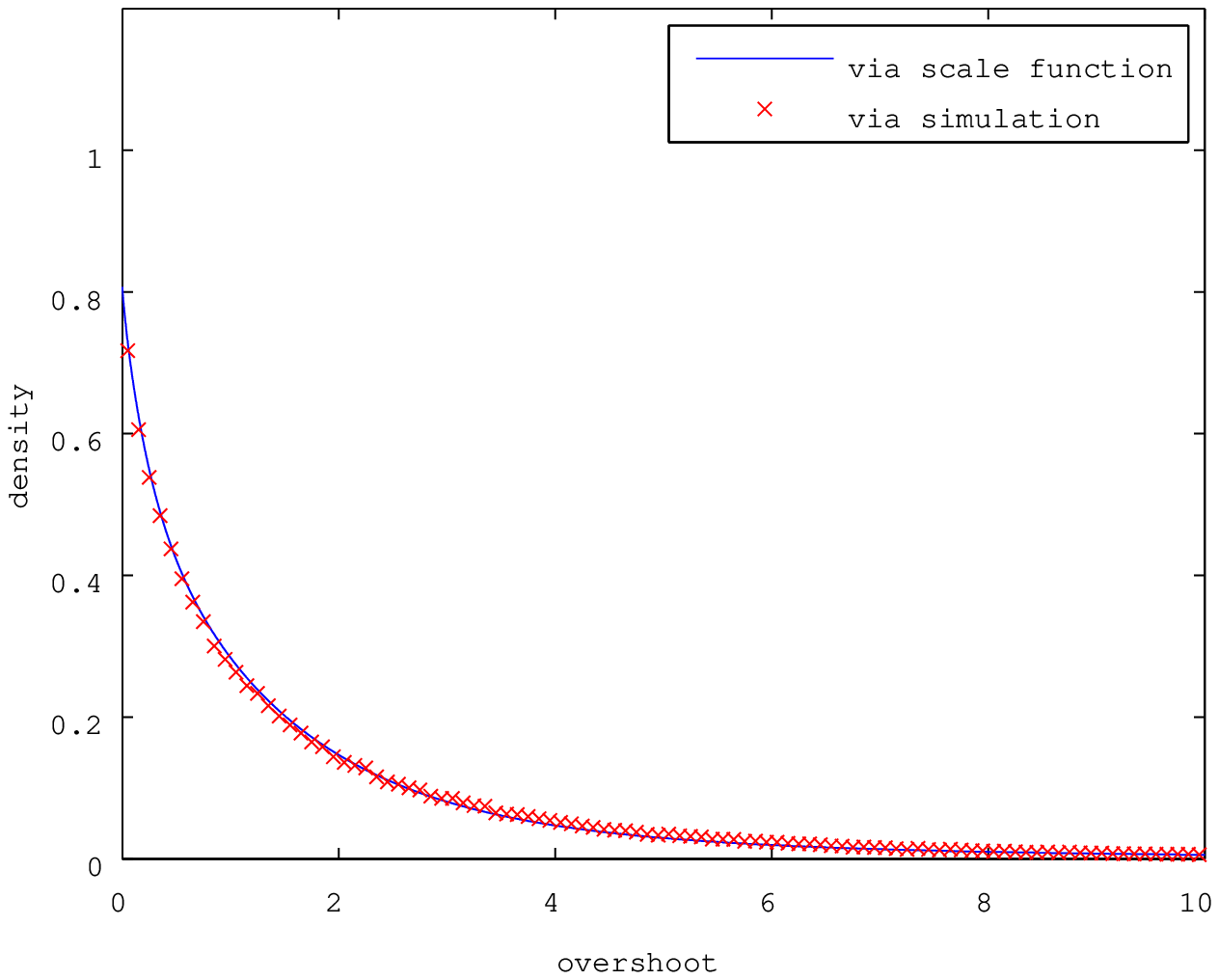} \\
Weibull(0.6,0.665) with $\sigma =1$ & Weibull(0.6,0.665) with $\sigma =0$ \vspace{0.3cm} \\
\includegraphics[scale=0.55]{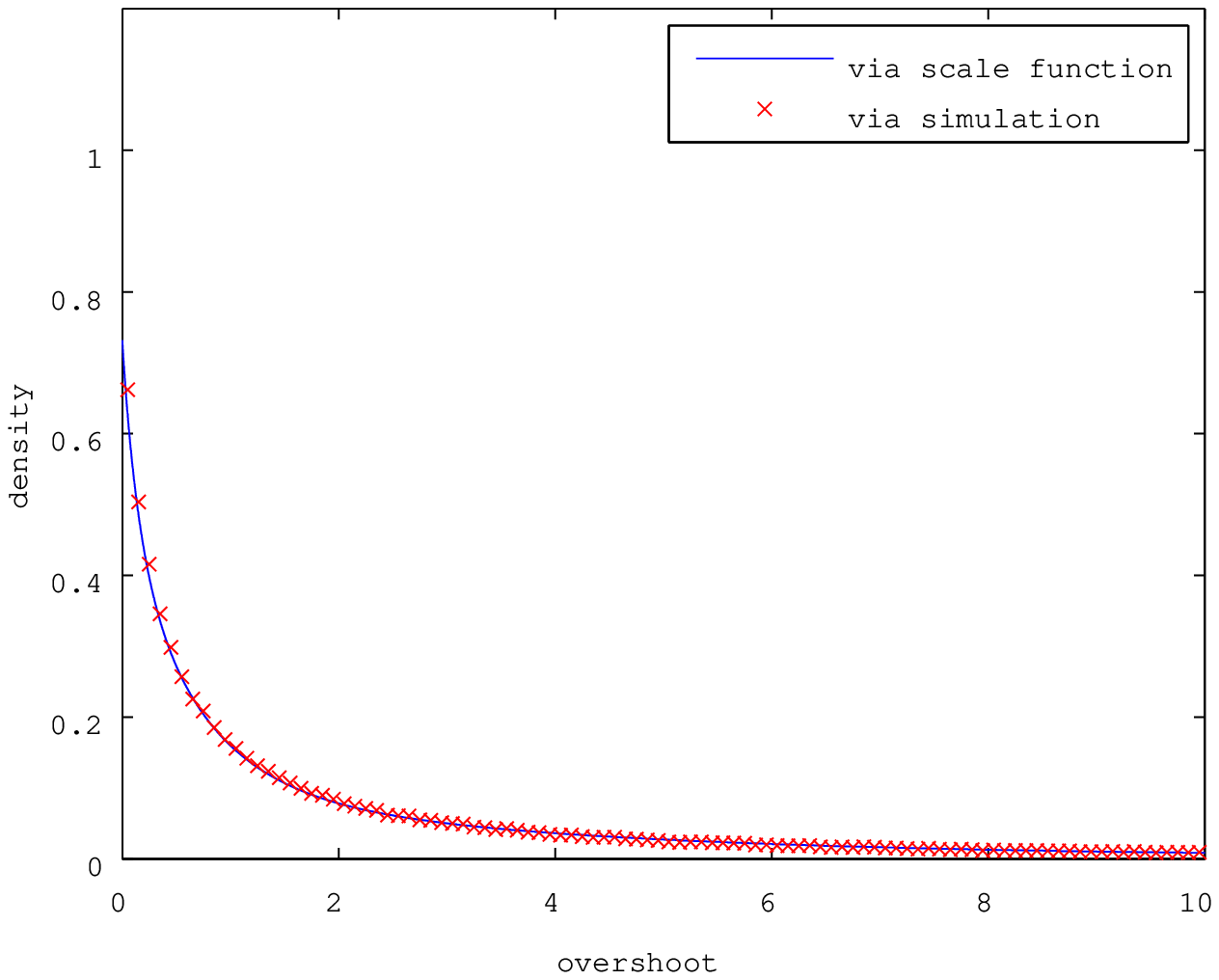}  & \includegraphics[scale=0.55]{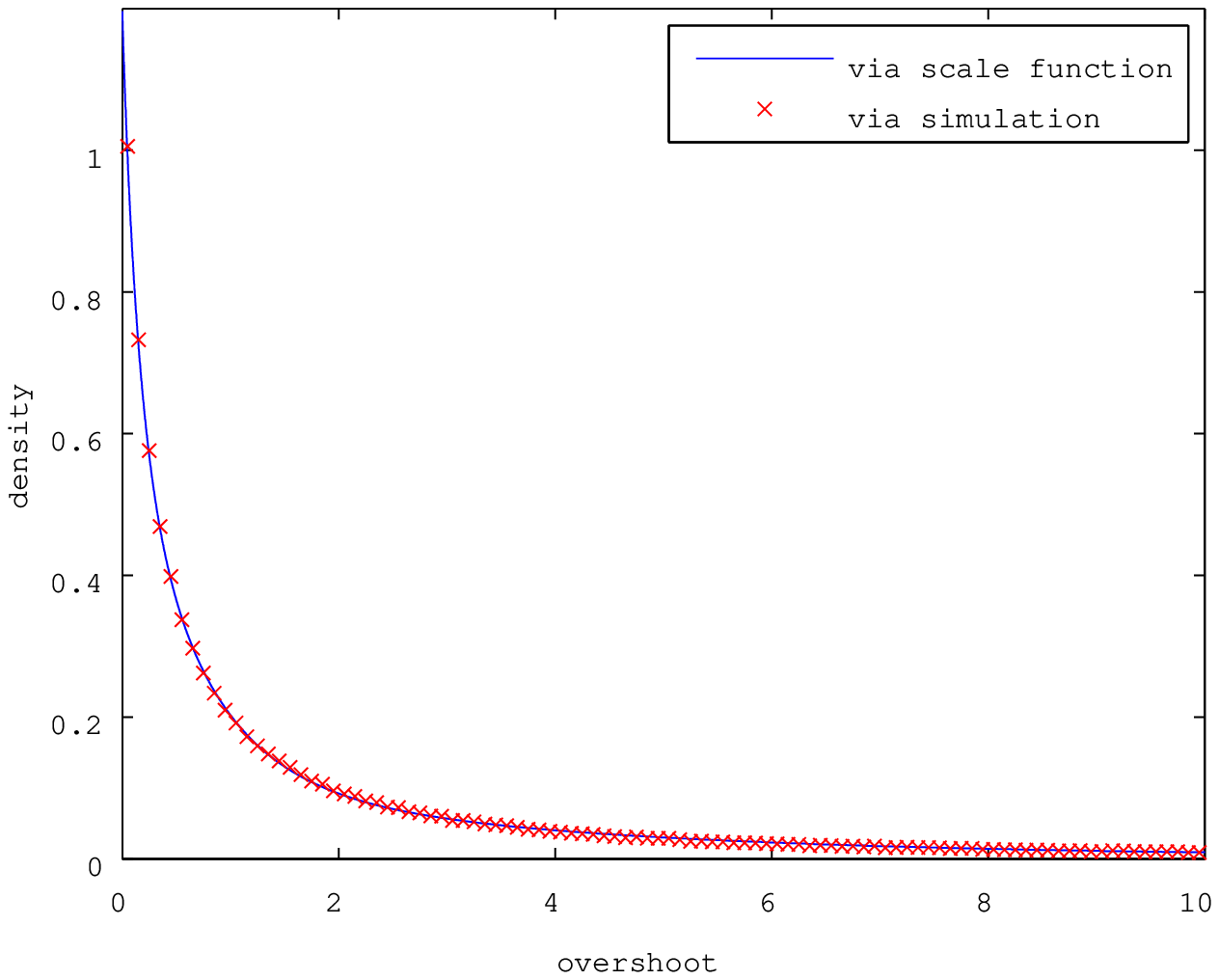} \\
Pareto(1.2,5) with $\sigma =1$ & Pareto(1.2,5) with $\sigma =0$ \vspace{0.3cm} \\
\end{tabular}
\end{minipage}
\caption{Computation of the overshoot density $\E^x [ e^{-q \tau_0^-}; -X_{\tau_0^-} \in \diff a, \tau_0^- < \infty]$. The solid lines indicate the fitted density functions and red marks indicate the values obtained from simulation.}
\label{fig:overshoot}
\end{center}
\end{figure}

\begin{figure}[htbp]
\begin{center}
\begin{minipage}{1.0\textwidth}
\centering
\begin{tabular}{cc}
\includegraphics[scale=0.55]{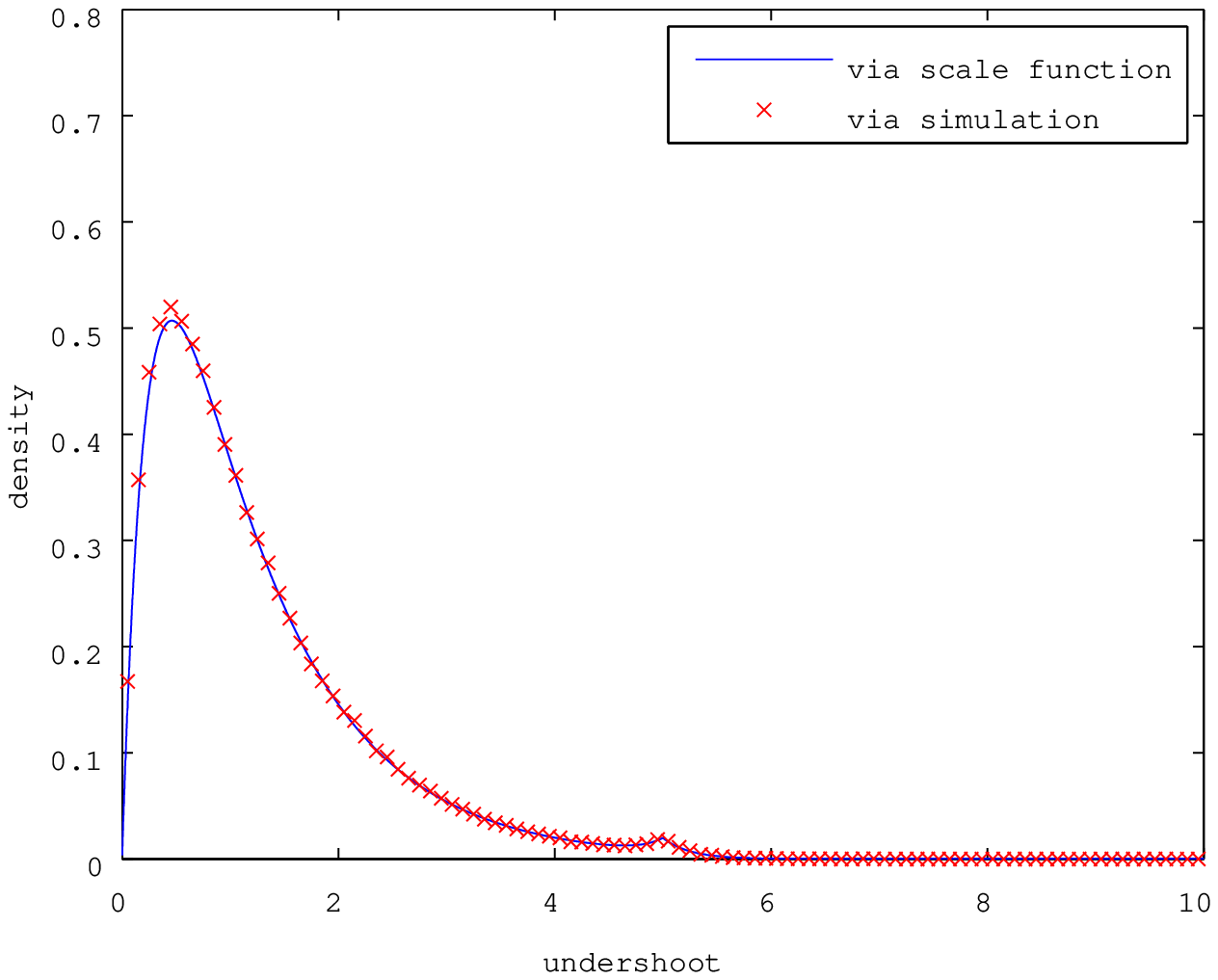}  & \includegraphics[scale=0.55]{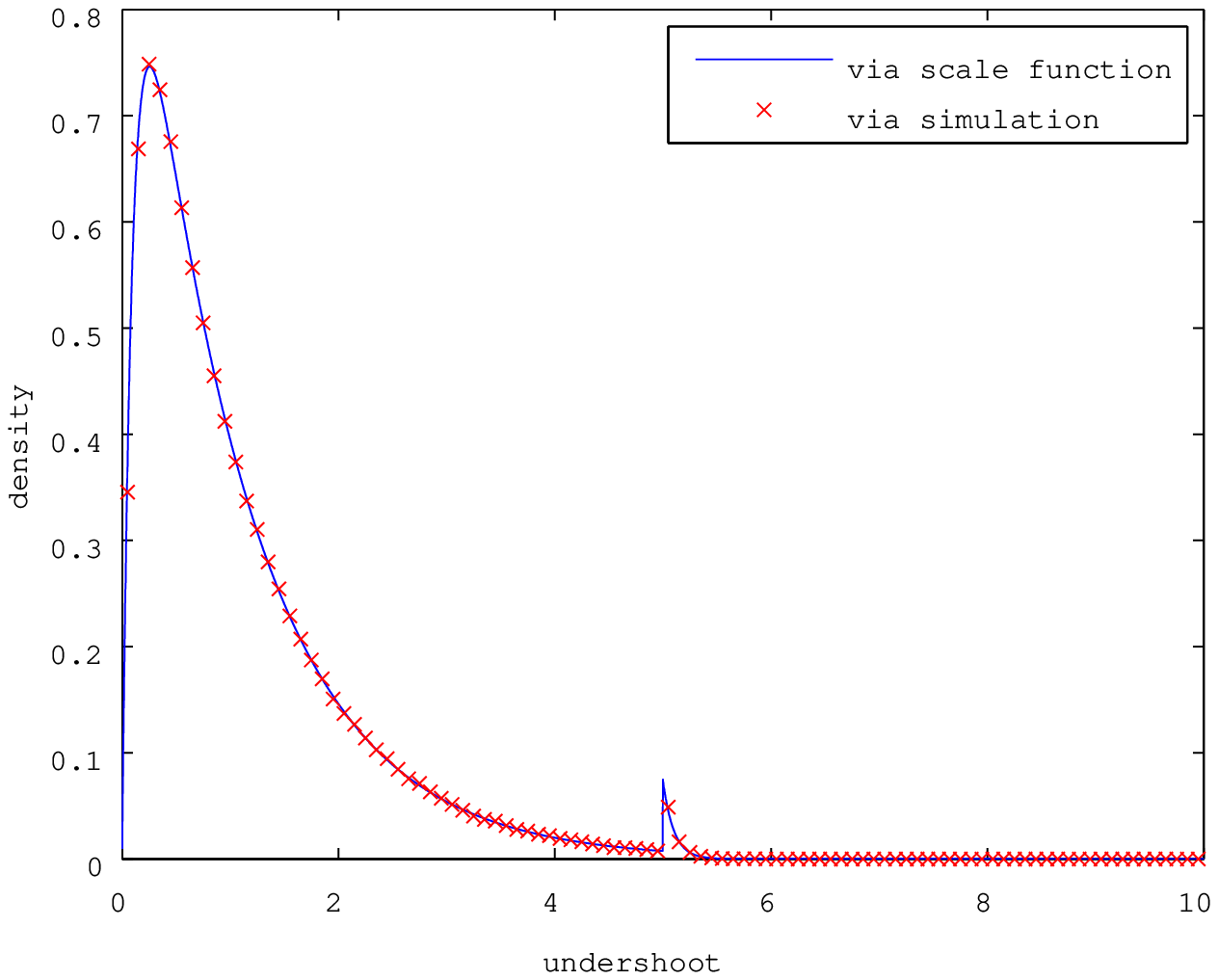} \\
Exp(1)  with $\sigma =1$ & Exp(1)  with $\sigma =0$ \vspace{0.3cm} \\
\includegraphics[scale=0.55]{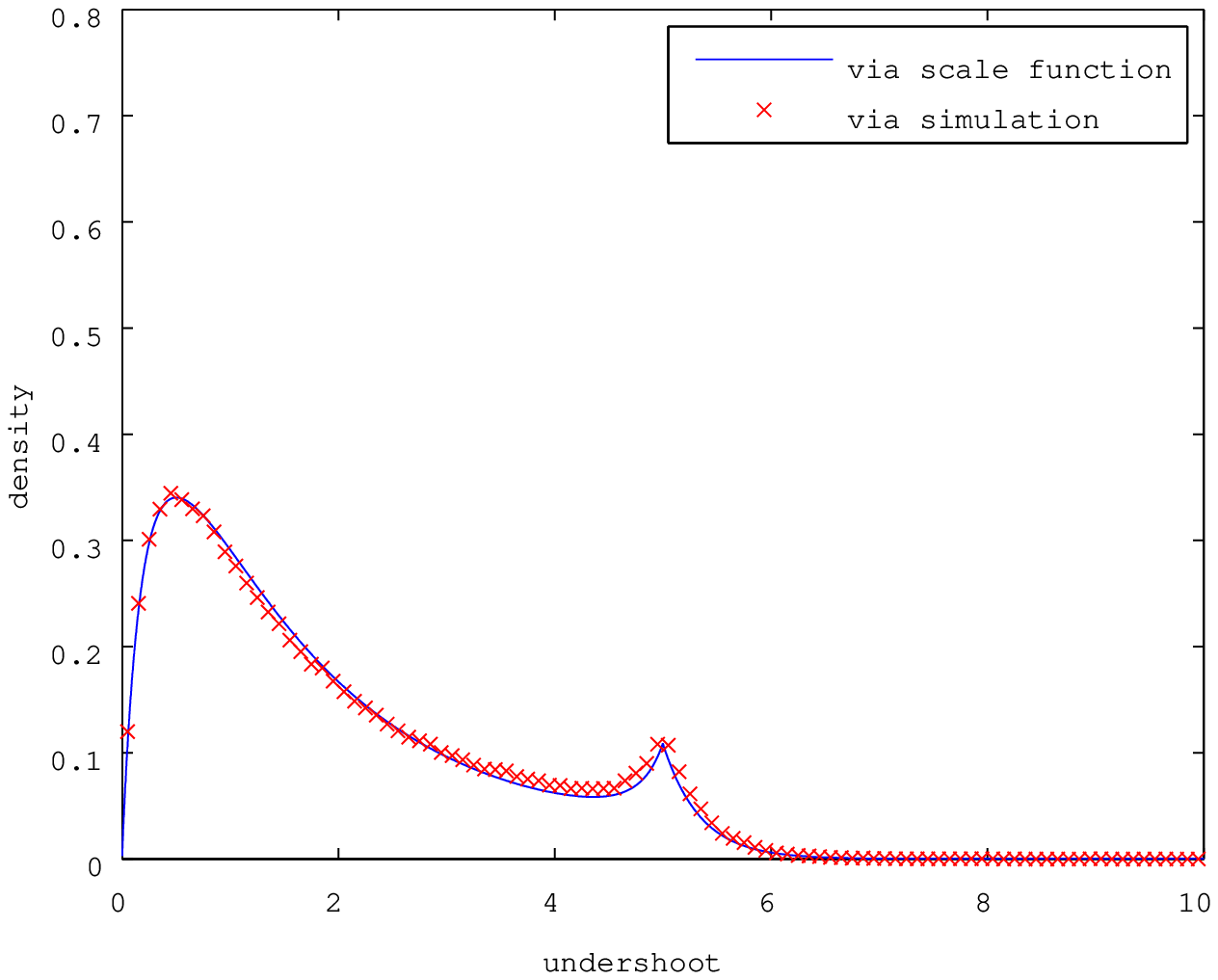}  & \includegraphics[scale=0.55]{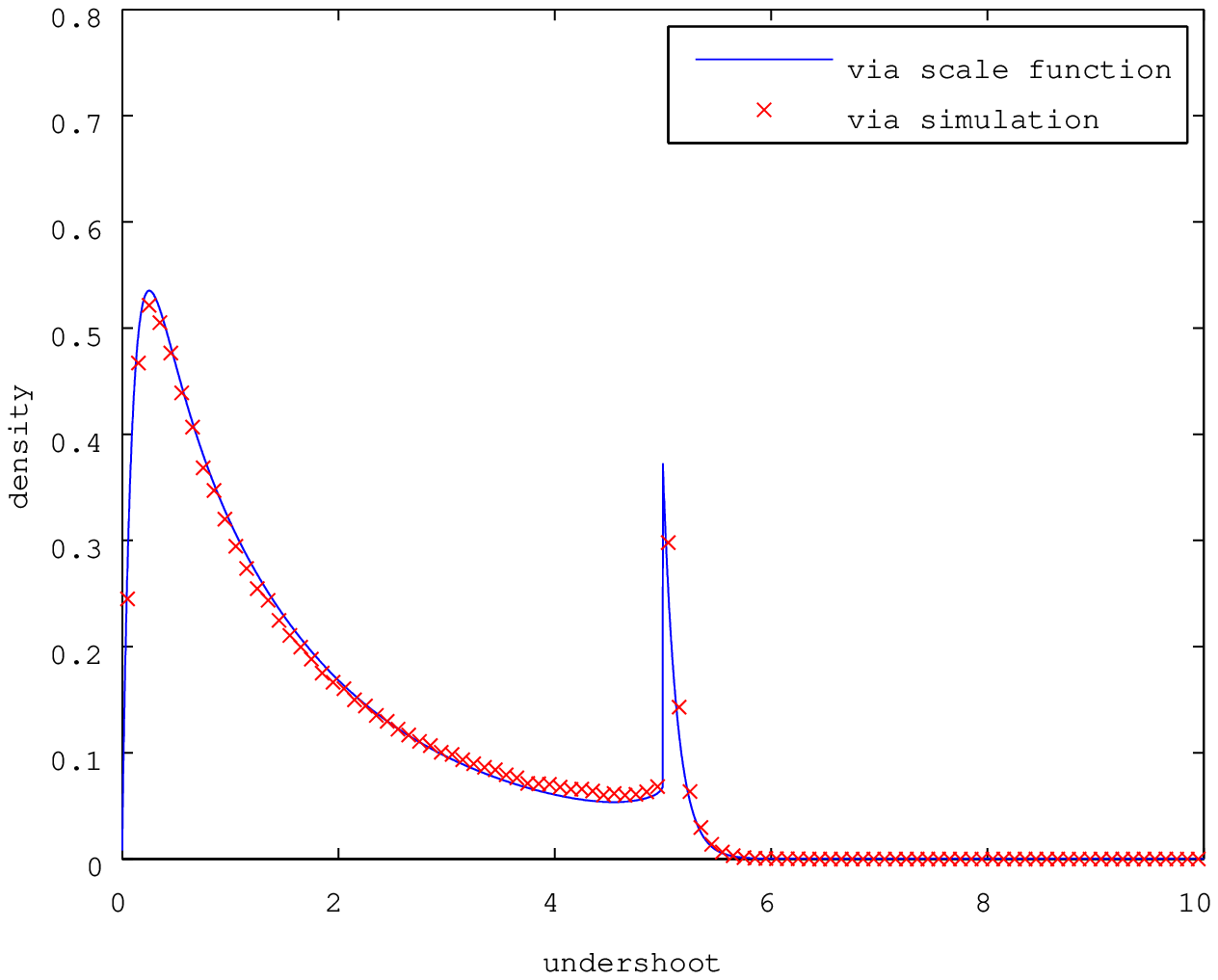} \\
Weibull(0.6,0.665) with $\sigma =1$ & Weibull(0.6,0.665) with $\sigma =0$ \vspace{0.3cm} \\
\includegraphics[scale=0.55]{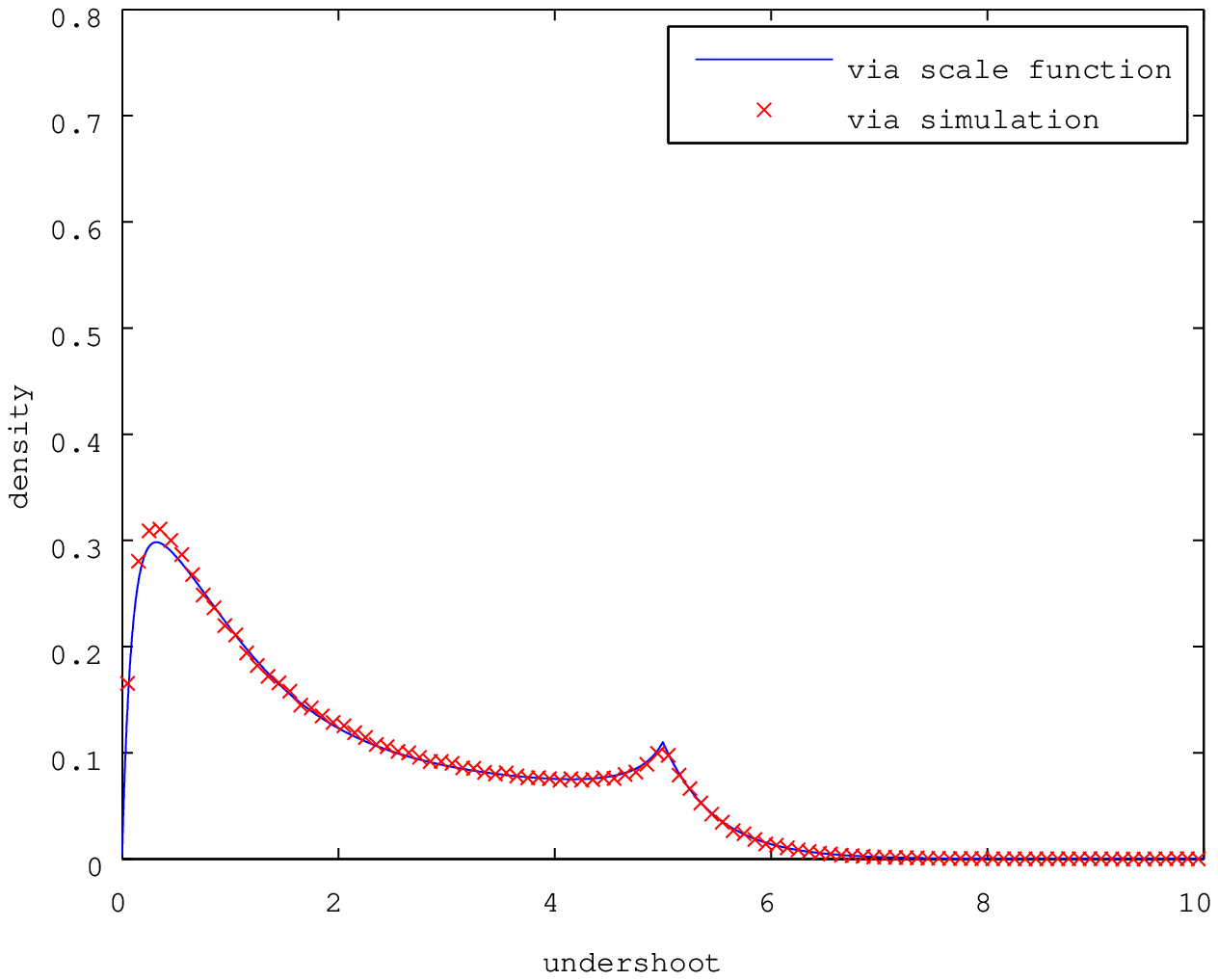}  & \includegraphics[scale=0.55]{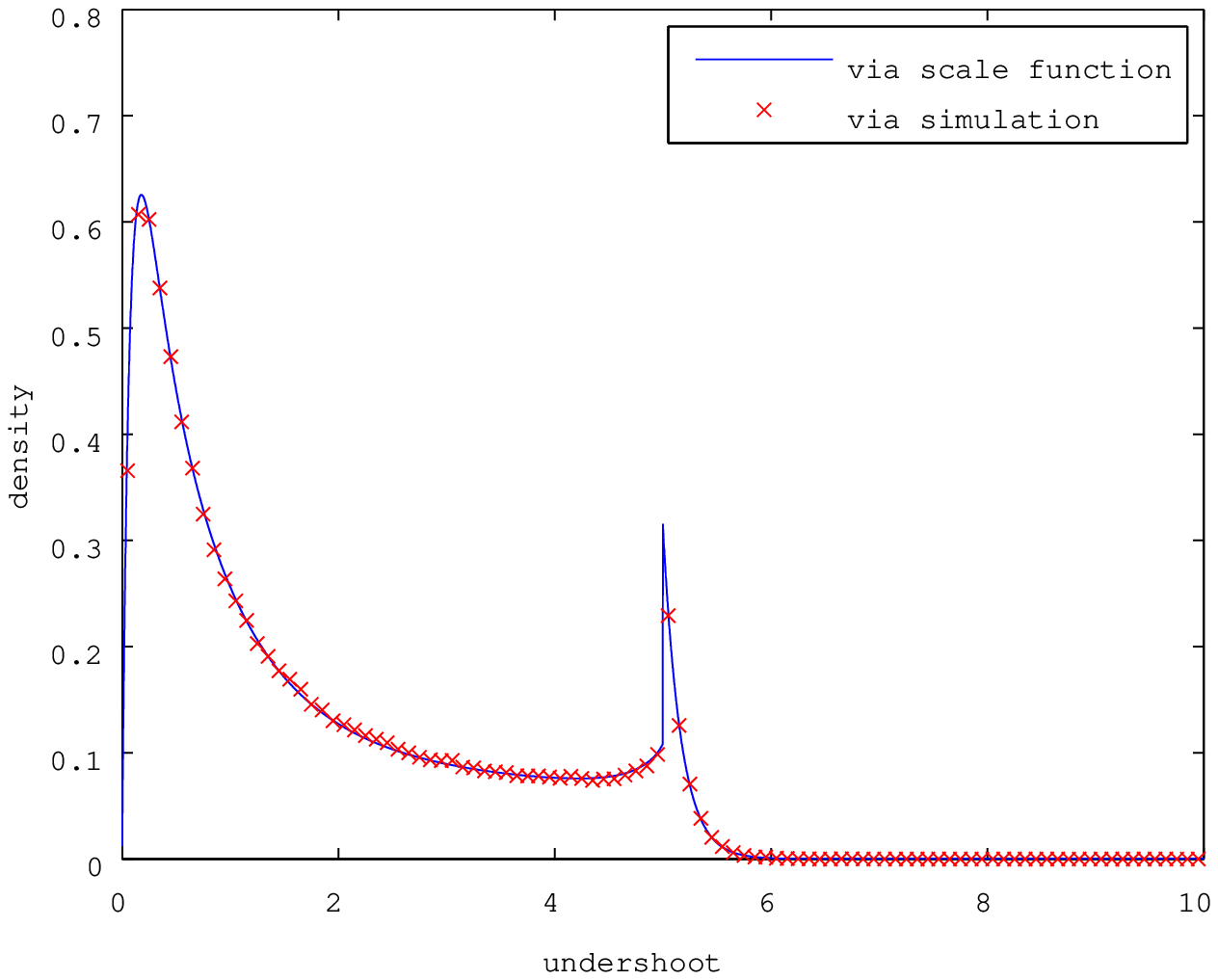} \\
Pareto(1.2,5) with $\sigma =1$ & Pareto(1.2,5) with $\sigma =0$ \vspace{0.3cm} \\
\end{tabular}
\end{minipage}
\caption{Computation of the undershoot density $\E^x [ e^{-q \tau_0^-};X_{\tau_{0}^--} \in \diff b, \tau_0^- < \infty ]$. The solid lines indicate the fitted density functions and red marks indicate the values obtained from simulation.}
\label{fig:undershoot}
\end{center}
\end{figure}

Figures \ref{fig:overshoot} and \ref{fig:undershoot} show the results for the cases $\sigma = 1$ and $\sigma = 0$ with common parameters $x=5$, $\mu = 1$, $\lambda = 10$ and $q = 0.05$. From the results on the exponential case, we can confirm that the Monte Carlo simulated results are accurate.  In Figure \ref{fig:undershoot}, the density has a jump at the initial position $x=5$ for the case $\sigma = 0$ (while it is continuous for the case $\sigma = 1$) due to the fact that $x=5$ is irregular  for $(-\infty,5)$ (see Definition 6.4 of \cite{Kyprianou_2006}).  As can be seen from these figures, the approximation accurately captures the overshoot/undershoot densities for $X^{(\textrm{weibull})}$  and $X^{(\textrm{pareto})}$.  The spike at the initial position in Figure \ref{fig:undershoot} is precisely realized thanks to the closed-form expression \eqref{overshoot_undershoot_density}; this would be difficult to realize if the scale function is approximated via numerical Laplace inversion.

\section{Concluding Remarks} \label{section_concluding_remarks}

In this paper, we studied and evaluated the performance of the phase-type fitting approach (motivated by \cite{Egami_Yamazaki_2010_2}) for computing the Gerber-Shiu function for the spectrally negative \lev process.  The method is overall accurate and is powerful in that it can obtain an approximation in a closed form.

While this paper focused on the case of a spectrally negative \lev process, the proposed method can easily be generalized.  By the compensation formula, the decomposition \eqref{gerber-shiu_in_terms_of_resolvent} holds for a more general class of stochastic processes. Therefore, by simply replacing the resolvent measure $r^{(q)}$, the Gerber-Shiu function can be computed in the same way.  
 
 Thanks to the recent developments of the fluctuation theories, the resolvents of various extensions of the spectrally negative \lev process are now available in terms of the scale function.  Here, we list several known examples of the resolvent. 

 \begin{enumerate}
 \item  The fluctuation theories of  \emph{reflected spectrally negative \lev processes} are well-developed. In the optimal dividend problem, where one wants to maximize the expected net present value of dividends until ruin, it is in many cases shown to be optimal to reflect the surplus process at a suitable boundary.  It is therefore of interest to investigate the Gerber-Shiu function of the reflected process to evaluate the risk of an dividend paying company.  As given in \cite{Pistorius_2004}, the resolvent requires the derivative or the integral of the scale function, depending on whether the reflection barrier is upper or lower.   For the doubly reflected case with both upper and lower barriers, see \cite{Pistorius_2003}.
\item As a variant of the reflected process, the \emph{refracted spectrally negative \lev process} of \cite{Kyprianou_Loeffen} changes its drift by $\delta > 0$ whenever it is above a threshold $b$ -- it is the unique strong solution $U$ to the stochastic differential equation
\begin{align*}
\diff U_t = \diff X_t - \delta 1_{\{ U_t > b \}} \diff t, \quad t \geq 0.
\end{align*}
In insurance, this can be used to model the surplus of a dividend paying company when the dividend rate must be bounded from above by $\delta$ (see \cite{Kyprianou_Loeffen_Perez}).  The resolvent is given in \cite{Kyprianou_Loeffen}.  The resolvents for the cases with additional classical reflection have recently been obtained by \cite{Perez_Yamazaki_2017_2, Perez_Yamazaki_2017}. 

\item  Given two levels $s$ and $S$, the classical $(s,S)$-policy controls the process by pushing the process to $S$ immediately when it goes above or below $s$.  It is, under a suitable condition, an optimal strategy in the optimal dividend problem in the presence of a fixed cost (see \cite{Bayraktar_2013, Loeffen_2008_2}).  The resolvent is obtained in \cite{Yamazaki_2016_2}.  For its two-sided cases with four-parameter $(d,D,U,u)$ policy, see \cite{Yamazaki_2016}.
 \end{enumerate}
 
In these examples, the resolvent  $r^{(q)}$ is written in terms of the scale function, and hence as in the cases considered in this paper, those for the phase-type case can be written as  (a linear combination of) exponential forms. Consequently, the Gerber-Shiu function can be analytically computed in the same way. In view of the results obtained in this paper, the phase-type error is minimal and the same procedure is expected to give an accurate approximation of the Gerber-Shiu function.


\vspace{15pt}
{\leftskip 8.5cm \parindent 0mm
Kazutoshi Yamazaki \\
Department of Mathematics \\
Faculty of Engineering Science \\
Kansai University \\
3-3-35 Yamate-cho, Suita-shi, \\
Osaka 564-8680, Japan \\
E-mail: \texttt{kyamazak@kansai-u.ac.jp}
\par}
\end{document}